\documentclass[%
superscriptaddress,
 amsmath,floatfix,amssymb,
 aps,
onecolumn
]{revtex4-1}

\usepackage{graphicx}
\usepackage{amssymb}
\usepackage{float}
\usepackage{amsmath}
\usepackage{verbatim}
\usepackage{graphicx}
\usepackage[toc,page]{appendix}
\usepackage{braket}
\usepackage{amsmath}
\usepackage{siunitx,esvect}
\usepackage{siunitx}
\usepackage{textcomp}
\usepackage{gensymb}
\usepackage{appendix}
\usepackage{braket}
\usepackage{upgreek}
\usepackage[breaklinks=true,colorlinks=true,linkcolor=blue,urlcolor=blue,citecolor=blue]{hyperref}
\usepackage{dcolumn}
\usepackage{bm}
\usepackage{natbib}
\usepackage{multirow}

\setcitestyle{super}

\begin{document}

\title{Thermo-mechanical design for a miniaturized quantum light source on board the SpooQy-1 CubeSat}
\author{Huai Ying Lim\footnote{Email address: huaiying@speqtral.space.}}%
\affiliation{%
 Centre for Quantum Technologies, National University of Singapore, 3 Science Drive 2, S117543, Singapore\\
}%
\author{Tom Vergoossen}%
\affiliation{%
 Centre for Quantum Technologies, National University of Singapore, 3 Science Drive 2, S117543, Singapore\\
}%
\author{Robert Bedington}%
\affiliation{%
 Centre for Quantum Technologies, National University of Singapore, 3 Science Drive 2, S117543, Singapore\\
}%
\author{Xueliang Bai}%
\affiliation{%
 Centre for Quantum Technologies, National University of Singapore, 3 Science Drive 2, S117543, Singapore\\
}%
\author{Aitor Villar}
\affiliation{%
 Centre for Quantum Technologies, National University of Singapore, 3 Science Drive 2, S117543, Singapore\\
}%
\author{Alexander Lohrmann}%
\affiliation{%
 Centre for Quantum Technologies, National University of Singapore, 3 Science Drive 2, S117543, Singapore\\
}%
\author{Nguyen Hong Nhung}%
\affiliation{%
 Centre for Quantum Technologies, National University of Singapore, 3 Science Drive 2, S117543, Singapore\\
}%

\author{Simon Barraclough}%
\affiliation{University of New South Wales Canberra, School of Engineering and Information Technology, Canberra, Australia\\
}
\author{Jai Vennik}%
\affiliation{University of New South Wales Canberra, School of Engineering and Information Technology, Canberra, Australia\\
}
\author{Douglas Griffin}%
\affiliation{University of New South Wales Canberra, School of Engineering and Information Technology, Canberra, Australia\\
}

\author{Alexander Ling}
\affiliation{%
 Centre for Quantum Technologies, National University of Singapore, 3 Science Drive 2, S117543, Singapore\\
}%
\affiliation{Physics Department, National University of Singapore, 2 Science Drive 3, S117542, Singapore}

\date{\today}

\begin{abstract}
This paper presents the thermo-mechanical design of the quantum light source on board SpooQy-1, a 3U CubeSat that was deployed from the International Space Station on 17th June 2019. SpooQy-1 is a technology demonstrator for space-based quantum networks. The on-board light source generates and detects polarization-entangled photon pairs to validate its in-orbit performance. Entangled photons are generated using spontaneous parametric down-conversion (SPDC) necessitating stringent dimensional stability and temperature requirements. Under laboratory conditions these requirements are routinely met using off-the-shelf laboratory mounts and alignment mechanisms. However, when facing harsh environments such as the vibration during rocket launch or temperature changes due to fluctuating illumination conditions, custom thermo-mechanical solutions are required. In this work, the development and in-orbit demonstration of an isostatic payload mount is discussed. This mounting approach enables future space missions with quantum instruments on resource-constrained CubeSat platforms with limited thermal control capabilities. 
\end{abstract}
\maketitle

\section{Introduction}

SpooQy-1 is a 3U CubeSat developed at the Centre for Quantum Technologies (CQT). The satellite was launched in April 2019 to the International Space Station (ISS) on board a Cygnus CRS-11 cargo mission and deployed into orbit from the ISS on the 17th of June 2019. SpooQy-1 is a technology demonstration mission of a miniaturized quantum light source which generates and detects polarization-entangled photon pairs in a low-Earth orbit (LEO). In this paper, we report a mechanical design and its in-orbit performance including flexure stages for optics mountings~\cite{tang2018towards} and an isostatic payload mounting structure.

The SpooQy-1 satellite is a pathfinder mission towards the goal of a global quantum internet which can enable applications such as Quantum Key Distribution (QKD). QKD, the most mature of quantum network applications, is a symmetric key distribution method achieved by manipulating the quantum states found in various degrees of freedom of light particles, where a correlated pair of secret keys (a random string of 1’s and 0’s) can be obtained securely between two distant parties. The security of QKD relies on the no-cloning quantum theorem~\cite{wootters1982single}, which states that it is impossible to generate an identical copy of an unknown quantum state.

Currently, QKD is seeing early deployment for securing day-to-day infrastructure and efforts are lined up~\cite{polnik2020scheduling} towards spanning continental distances with QKD links. In this regard, high-altitude nodes (e.g., drones, satellites) play an important role, since traditional optical links based on fiber or terrestrial free-space communications are fundamentally limited to few hundred kilometers. Recently, the 630kg Chinese satellite Micius demonstrated various quantum communication primitives from space including space-to-ground quantum key distribution~\cite{liao2017satellite}. Miniaturizing technologies to increase cost-effectiveness of QKD networks and facilitate widespread adoption is a topic of current research~\cite{kerstel2018nanobob,morong2012quantum,oi2017cubesat}.

The Centre for Quantum Technologies started the Small Photon Entangling Quantum System (SPEQS) program in 2012. This envisions a space-based quantum network using a constellation of smaller spacecraft such as nano-satellites. Following an iterative approach~\cite{morong2012quantum,tang2014near,tang2016generation,tang2016photon,bedington2016nanosatellite}, an entangled photon source was space-qualified and installed in the SpooQy-1 CubeSat. This light source is capable of generating polarization-entangled photons, and the quality of the entanglement was tested in-situ on the spacecraft.

Entangled photon sources based on spontaneous parametric down-conversion have stringent dimensional stability and temperature requirements. This is straightforward under laboratory conditions. However, launch conditions and operation in space impose a challenging environmental envelope. For instance, in an orbit following the International Space Station track, the internal temperature of the satellite fluctuates from approximately \SI{0}{\degree C} to \SI{30}{\degree C}. In terms of dimensional stability, changes in temperature introduce optical misalignment in the quantum light source affecting entanglement generation. Active thermal control capabilities are typically out of the size, weight and power (SWaP) budget in nano-satellite missions~\cite{clements2016nanosatellite,oi2017nanosatellites,naughton2019design}. For this reason, thermally isolating the instrument from the satellite bus such that the temperature changes are reduced is necessary in order to meet mission objectives.

In this work, the development of an isostatic payload mounting approach on a nano-satellite is discussed. This mounting approach paves the way for future resource-constrained missions with quantum instruments, where tackling temperature changes or gradients might affect the instrument performance.

\section{SpooQy-1 System Design}

\begin{figure}[h]
	\centering
	\includegraphics{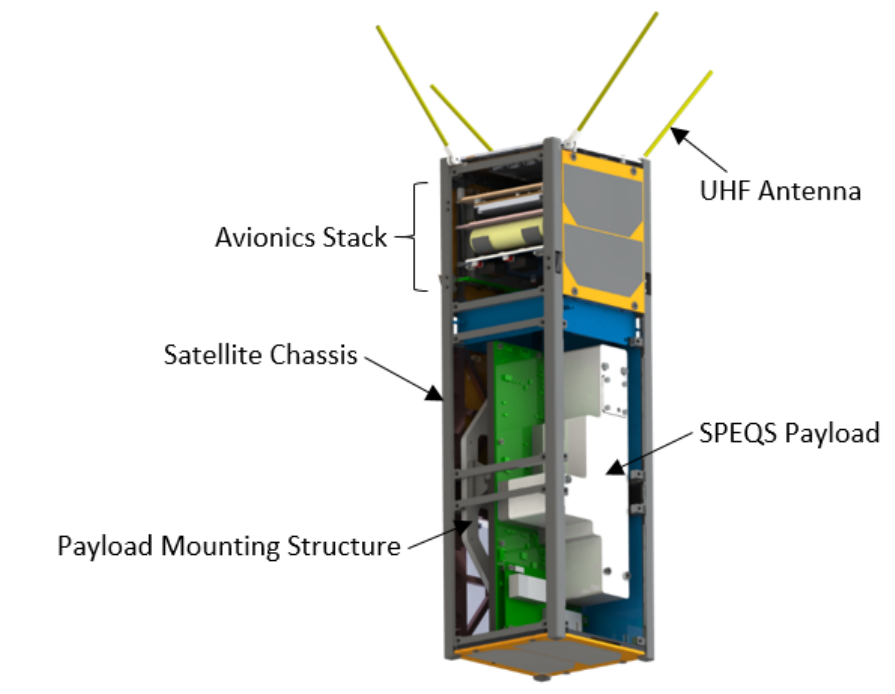}
	\caption{SpooQy-1 CAD model with external panels removed.}
	\label{fig:set1}
\end{figure}

SpooQy-1 is a 3-unit (3U) CubeSat measuring \SI{340}{} x \SI{100}{} x \SI{100}{\milli\metre}. 1U is allocated for the satellite avionics stack and 2U is for the SPEQS payload, as illustrated in Fig.~\ref{fig:set1}. The avionics stack is comprised of the standard GomX platform structural parts and subsystems first demonstrated on the GomX-3 mission launched in 2015 from the ISS. The satellite configuration is shown in Fig.~\ref{fig:set1}.

The avionics stack consists of the communication subsystem, the on-board computer (OBC), as well as the electrical power subsystem (EPS) along with the battery pack. The system configuration of SpooQy-1 is summarized in Table~\ref{tab:param}. The system design of SpooQy-1 as well as the ground station configuration have been discussed in detail in previous work~\cite{bai2018validating}.
\begin{table}
\begin{center}
\begin{tabular}{|c|c|} 
\hline
Bus & 3U CubeSat\\
\hline
Size & \SI{100}{} x \SI{100} x \SI{340}{\milli\metre}\\
\hline
Mass & \SI{2.6}{\kilogram}\\
\hline
RF Communications & UHF (\SI{436.2}{\mega\hertz}) in downlink and uplink\\
\hline
Power & \SI{3.85}{\watt} (max consumption)\\
\hline
Battery & \SI{38}{\watt hr} Li-ion\\
\hline
Solar Panels & GaAs cells\\
\hline
Flight Computer & 32-bit AVR processor\\
\hline
Attitude Control & Passive\\
\hline
\end{tabular}
\caption{SpooQy-1 System Configuration.}
\label{tab:param}
\end{center}
\end{table}

\section{Quantum Light Source Payload Design}
\begin{figure}[t]
	\centering
	\includegraphics{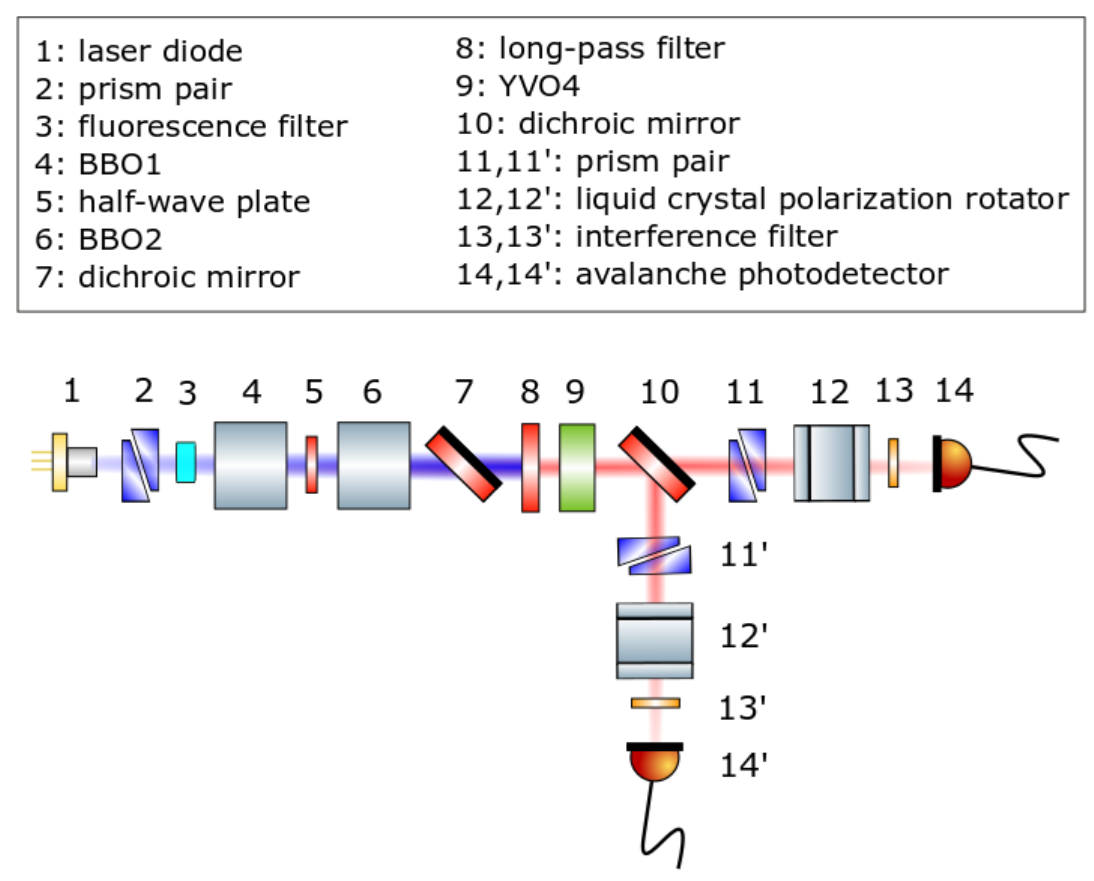}
	\caption{SPEQS optical layout.}
	\label{fig:set2}
\end{figure}

A few configurations of SPEQS based on different crystal arrangements have been explored during the development of SpooQy-1~\cite{bedington2015small,durak2016next,bedington2016nanosatellite,villar2018experimental,lohrmann2018high}. The commonality in the different configurations is the use of nonlinear optical crystals whose birefringent properties are not much affected by temperature. Fig.~\ref{fig:set2} illustrates the final optical layout of the SPEQS payload on board SpooQy-1.

The most crucial optical components are the two BBO crystals – element ‘4’ and ‘6’ in the optical layout illustrated in Fig.~\ref{fig:set2}, in which the entangled photons are generated. These crystals need to be aligned with the optical path with a maximum tolerable angular deviation of \SI{100}{\micro\radian} with respect to the pump beam generated by the laser diode. The rest of the optics and opto-electronic components have less stringent alignment requirements, however the cumulative misalignment, measured in \SI{}{\micro\radian}, can result in poor system performance.

\section{Payload Mechanical Design Overview }
In order to facilitate the calibration of the BBO crystals with high precision, flexure stages have been designed and tested extensively prior to integration into the optical unit. Titanium was chosen as the material for the flexure stages mainly due to its high yield strength and low coefficient of thermal expansion (CTE). In order to minimize the effect of the CTE mismatch within the unit, the optical unit base which houses all the components illustrated in Fig.~\ref{fig:set2} is also made of titanium.

The main electronics board containing the driving circuits for the opto-electronic components, is placed directly underneath the optical unit and secured onto a titanium baseplate. During payload operation when the laser diode is operating, the heat dissipated can lead to an uneven temperature distribution across the unit. This temperature gradient could cause a misalignment of the BBO crystals due to any thermo-elastic deformations. To mitigate this, a thin-film heater is placed underneath the optical unit base at the section where the BBO crystals are located.

\begin{figure}[h]
	\centering
	\includegraphics{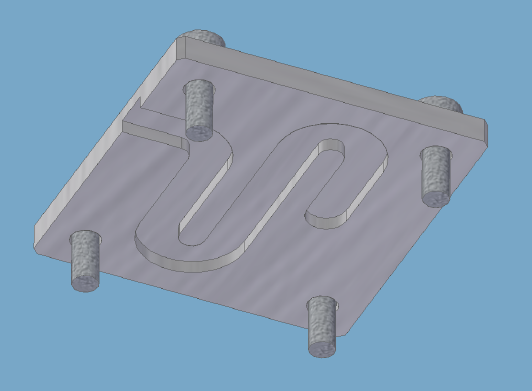}
	\caption{Air escape labyrinth path.}
	\label{fig:set3}
\end{figure}

\begin{figure}[h]
	\centering
	\includegraphics{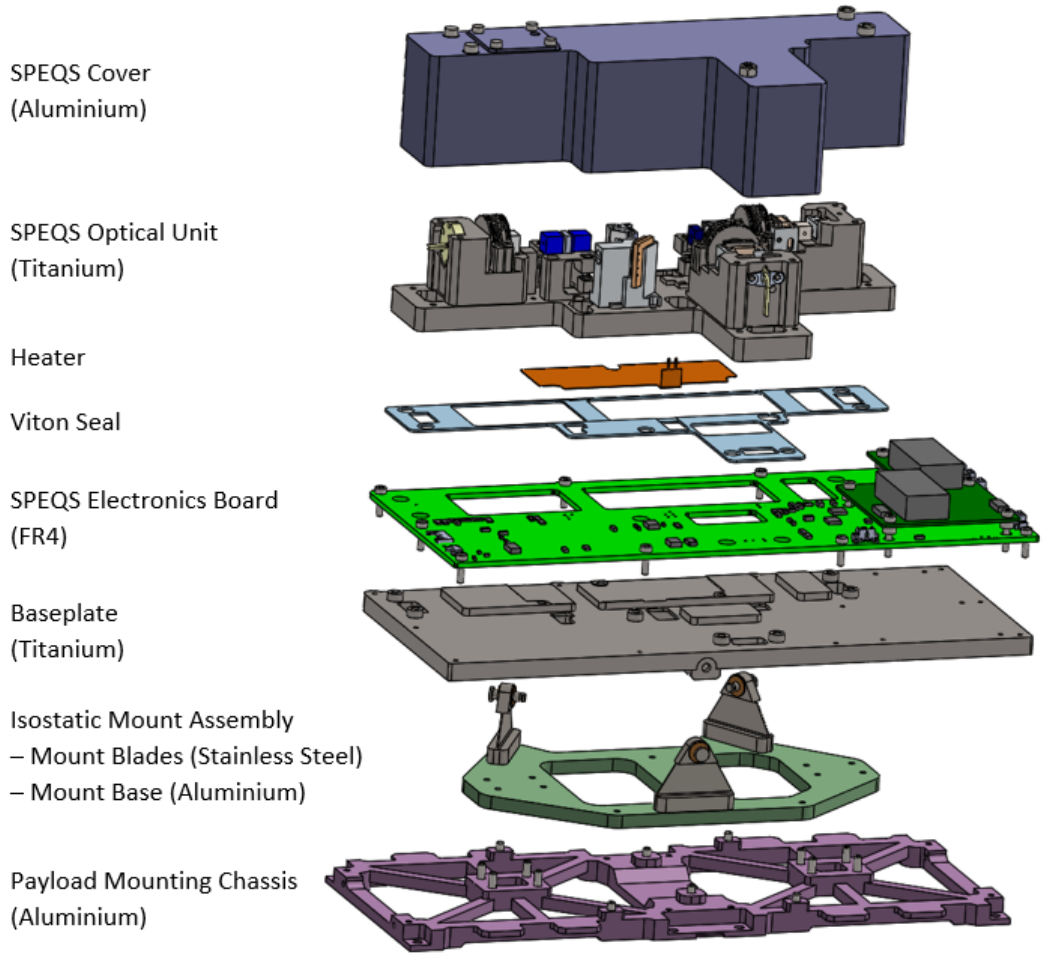}
	\caption{Overall payload configuration in exploded view.}
	\label{fig:set4}
\end{figure}

The optical unit needs to be light-tight to minimize stray light reaching the single photon detectors. Hence, it is enclosed by a black-anodized aluminium cover. To allow the optical unit to depressurize in space, without allowing light to enter the system, an air escape labyrinth path, illustrated in Fig.~\ref{fig:set3}, has also been included as part of the cover assembly.
\newpage
The overall payload configuration including the mounting structure is illustrated in Fig.~\ref{fig:set4}.

\section{Isostatic Payload Mounting Structure}
While the heater may help to reduce the thermal gradient across the unit due to internal heat dissipation, the SPEQS optical unit is also susceptible to external influences such as the overall change in satellite body temperature in orbit. Since the optical unit is assembled under room temperature at approximately \SI{22}{\degree C} to \SI{24}{\degree C}, this is then the ideal temperature range for SPEQS to operate without suffering misalignments due to thermo-elastically-induced deformations. This in turn places a constraint on the external thermal environment for SPEQS. In addition, the laser diode also has an operating temperature range of \SI{15}{\degree C} to \SI{28}{\degree C}, outside of which it may be damaged. Additional thermal consideration is thus required for SpooQy-1 to ensure a safe operating environment for the SPEQS payload.

Due to the limited SWaP available on a 3U CubeSat, no other active thermal control is implemented on SpooQy-1, besides the heater. To minimize temperature changes experienced by the optical unit, the payload is mounted on an isostatic mounting structure which provides thermal isolation between the payload and the satellite chassis.

\begin{figure}[h]
	\centering
	\includegraphics{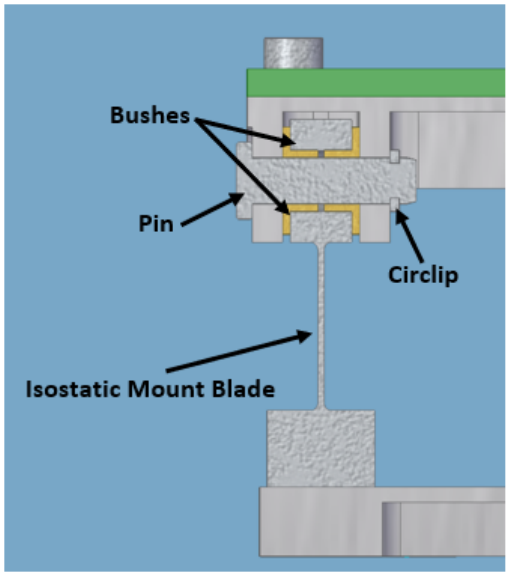}
	\caption{Isostatic mount configuration.}
	\label{fig:set5}
\end{figure}

The isostatic mount assembly consists of an aluminium base and three isostatic mounts that are set at angles of \SI{120}{\degree} apart with respect to each other, as shown in Fig.~\ref{fig:set4}. The isostatic mount consists of a mount blade acting as a flexing member of \SI{0.4}{\milli\metre} thickness, two bushes and a pin. Both the mount blade and the pin are made of stainless steel for its high yield strength and low thermal conductivity, whereas the bushes are made of phosphor bronze to provide frictionless motion between the mount blade and the baseplate onto which the SPEQS optical unit is mounted. A circlip is inserted at the end of the pin to secure it while allowing a rotational motion of the bushes. The relative expansion at the interface between the satellite and the payload is therefore taken up by the flexing member within the mount blade and the rotation of the bushing at all three interfaces between the isostatic mount and the baseplate. The 120° angle between the isostatic mounts provides the necessary mechanical support for the payload in vibration environments. The configuration of the isostatic mount is illustrated in Fig.~\ref{fig:set5}.

\section{Environmental Testing}
\begin{figure}[h]
	\centering
	\includegraphics{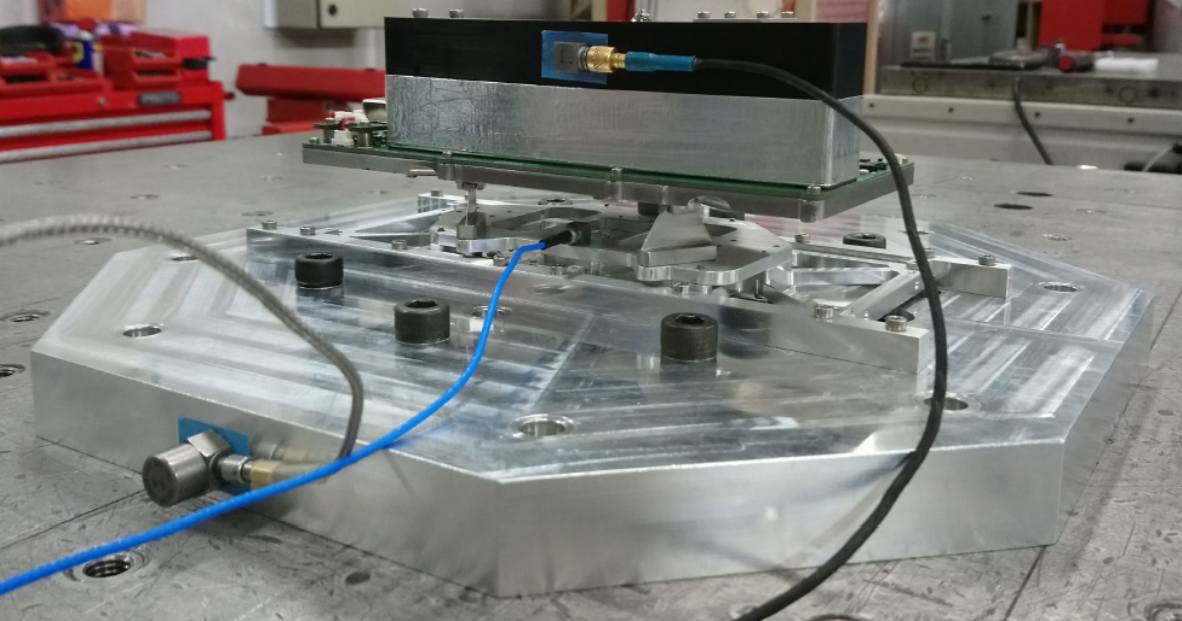}
	\caption{Random vibration test performed for the Qualification Model of the payload.}
	\label{fig:set6}
\end{figure}

The SpooQy-1 mission follows a traditional spaceflight model philosophy; hence it has gone through both Qualification Test (QT) and Acceptance Test (AT). Each test consisted of random vibration tests, thermal bake-out, and a thermal vacuum test. The random vibration test level was 10 g rms for the Qualification Test and \SI{7.1}{\gram rms} for the Acceptance Test. The test profile was consistent for all the tests which followed the requirement stated in the interface control document (ICD) from the Japan Aerospace Exploration Agency (JAXA)\cite{jaxaSpace}. Sine sweep tests performed before and after both QT and AT showed that the first eigenfrequencies for all three axes did not shift problematically. Functional tests confirmed the payload performance was comparable to pre-testing conditions, confirming that the isostatic mounting structure is structurally sound and can withstand the vibrations during launch. Fig.~\ref{fig:set6} shows the vibration test setup for the Qualification Model (QM) of the payload.

\begin{figure}[h]
	\centering
	\includegraphics{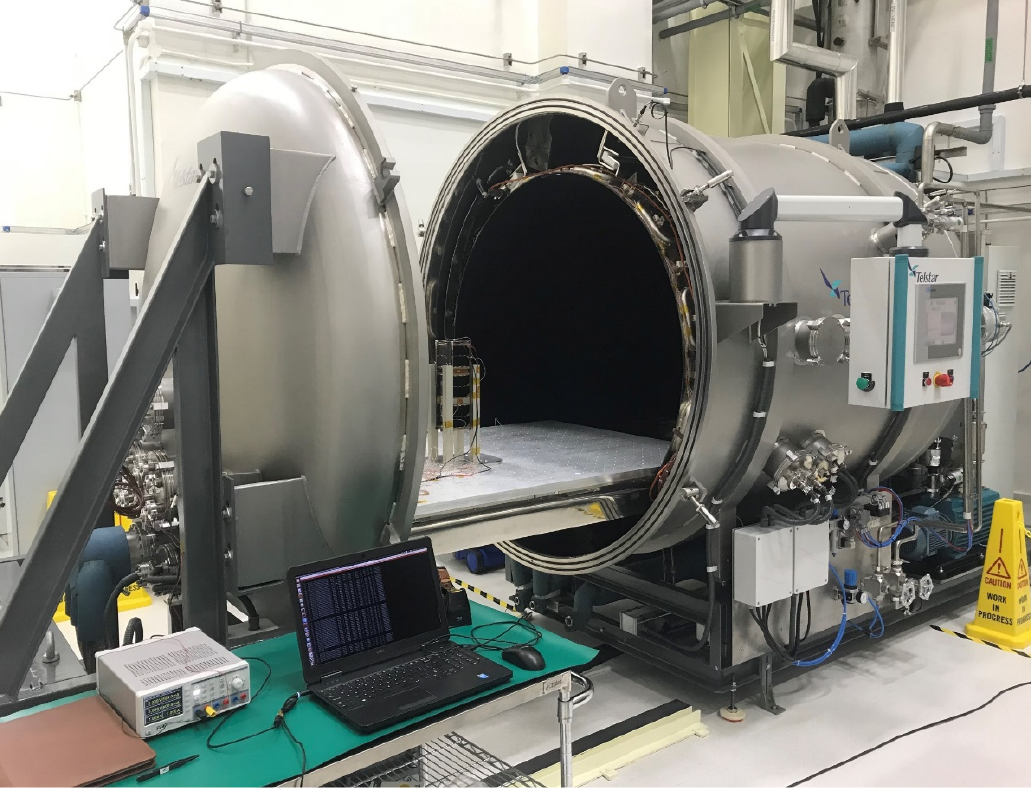}
	\caption{Thermal vacuum test performed for SpooQy-1 Flight Model.}
	\label{fig:set7}
\end{figure}

The temperature range for the thermal vacuum test was derived from thermal analysis results, with margins allocated based on guidelines from the European Cooperation for Space Standardization (ECSS)\cite{esaSpace}. The internal temperature of SpooQy-1 was expected to range from \SI{0}{\degree C} to \SI{30}{\degree C}. With a \SI{10}{\degree C} margin, the temperature range for the Qualification Test was determined to be from \SI{-10}{\degree C} to \SI{40}{\degree C}, whereas in the Acceptance Test, the satellite was thermally cycled from \SI{-5}{\degree C} to \SI{35}{\degree C}, i.e. \SI{5}{\degree C} temperature margin. Fig.~\ref{fig:set7} shows the test setup for the thermal vacuum test performed for the Flight Model.

The thermal vacuum test conducted for the QM consisted of 8 cycles in total, with 1 hour soaking at both ends to ensure thermal stabilization, while the test for the FM comprised only 2 thermal cycles. Functional tests performed after both QT and AT showed that all satellite subsystems, including the payload, functioned nominally.

The main performance metric for the payload is the entanglement quality of the entangled photons it can generate. This is assessed by the Clauser-Horne-Shimony-Holt (CHSH) parameter~\cite{clauser1969proposed}. Entanglement is achieved when a CHSH value of greater than 2 is obtained, and the theoretical limit of this parameter is approximately 2.82.

The CHSH parameter can be extracted from the visibility of the polarization correlation curves obtained from payload experiments. Before the Acceptance Test, a set of experiments was run and a CHSH parameter of \SI[separate-uncertainty = true]{2.63(7)}{} was recorded. After the thermal vacuum test, another set of experiments was run and the corresponding CHSH parameter was \SI[separate-uncertainty = true]{2.60(6)}{}. This showed that the SPEQS payload could still produce high quality entanglement after the Acceptance Test, concluding a successful space qualification campaign for SpooQy-1. More details on the payload data can be found in Appendix~\ref{appendix:payload-data}.

\section{In-orbit Data}
SpooQy-1 has been operational since its deployment from the ISS in June 2019. After the initial in-orbit commissioning phase, a few experiments were run on SPEQS, confirming that the opto-electronic components worked nominally. Housekeeping data was obtained and correlated to the data obtained from SPEQS.

\begin{figure}[h]
	\centering
	\includegraphics{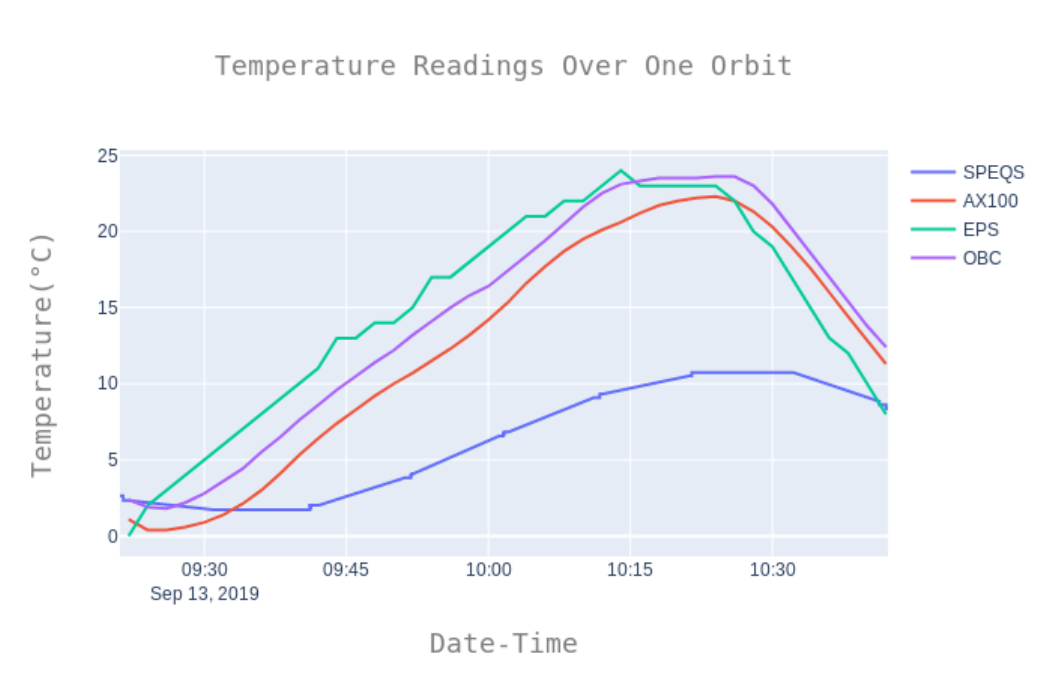}
	\caption{Temperature readings for SpooQy-1 subsystems.}
	\label{fig:set8}
\end{figure}
Fig.~\ref{fig:set8} shows the temperature readings of SpooQy-1 subsystems as well as the payload taken over an orbit, for a pass on 13th September 2019. From the figure, it can be seen that the temperature of the bus, which consists of EPS, OBC, and the radio transceiver, denoted as AX100, fluctuates between \SI{0}{\degree C} to \SI{25}{\degree C}, whereas the payload, denoted as SPEQS, experienced a much smaller temperature fluctuation in the range of \SI{2}{\degree C} to \SI{11}{\degree C}. A delay in temperature rise is also observed in the payload temperature profile when compared with the other subsystems. The smaller temperature fluctuation experienced by the payload as well as the delay in its temperature rise can be attributed to the thermal isolation provided by the isostatic mounting structure.

Within the first 6 months after deployment, several experiments were run, and data was obtained to evaluate the payload performance. Similarly, a set of polarization correlation curves were obtained and compared with the data taken after the Acceptance Test, as presented in Fig.~\ref{fig:appendix1}. The highest recorded visibilities were obtained from a satellite pass on 16th July 2019 and the corresponding CHSH parameter was \SI[separate-uncertainty = true]{2.60(7)}{}. This showed that the payload performance was comparable to pre-launch conditions, confirming that the isostatic mounting structure, as well as all other mechanical components within the optical unit were adequately designed for the SpooQy-1 mission.

\section{Conclusion and Outlook}
The generation of quantum entanglement on board SpooQy-1 was a challenging experiment due to the environmental conditions associated with a rocket launch and operations in a low-Earth orbit. The thermo-mechanical strategy of SPEQS proved suitable for operations in space and high-quality entanglement has been generated routinely since its deployment from the ISS on the 17th of June 2019. The mechanical elements developed through the SPEQS program, as well as the isostatic payload mounting approach can enable launching other scientific instruments on resource-constrained platforms. Specifically, the miniaturized quantum light source on board SpooQy-1 is a precursor towards demonstrating quantum networks from a CubeSat platform. An upcoming mission of the Centre for Quantum Technologies, SPEQTRE (formerly QKD QubeSat) will inherit many of the lessons of the SpooQy-1 design process.
\section{acknowledgments}
This program was supported by the National Research Foundation (Award No. NRF-CRP12-2013-02), Prime Minister’s Office of Singapore and the Ministry of Education, Singapore. The SPEQS payload was designed and built at the Centre for Quantum Technologies, National University of Singapore (NUS). The isostatic mounting structure was designed by the team at the University of New South Wales (UNSW), Canberra, with the detailed implementation done by CQT. 
\clearpage
\bibliographystyle{apsrev4-2}
\bibliography{bibliography}

\clearpage
\onecolumngrid
\begin{appendices}
\renewcommand\thefigure{\thesection.\arabic{figure}}
\renewcommand{\thefigure}{A\arabic{figure}}
\renewcommand{\theHfigure}{A\arabic{figure}}
\setcounter{figure}{0}
\section{Payload Data}
\label{appendix:payload-data}

\begin{figure}[h]
	\centering
	\includegraphics{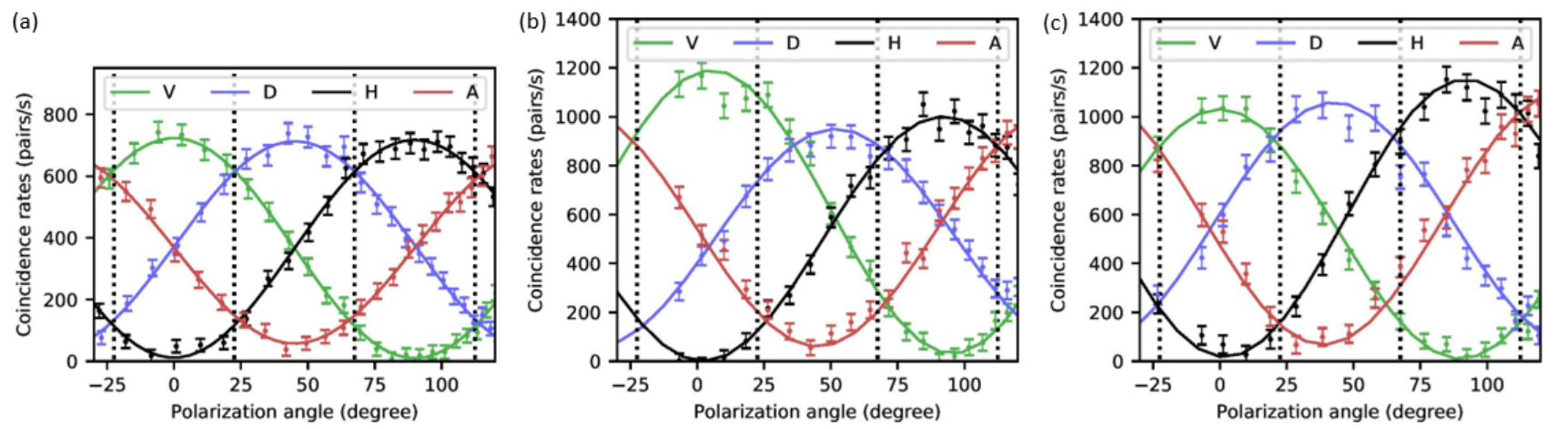}
	\caption{Experimental polarization correlation curves taken (a) before the Acceptance Test, (b) after the Acceptance Test, and (c) in orbit on 16th July 2019.}
	\label{fig:appendix1}
\end{figure}

The entanglement quality of the payload can be assessed by the Clauser-Horne-Shimony-Holt (CHSH) parameter, which can be extracted from the visibility of the polarization correlation curves obtained from the payload experimental runs. For each experimental run, a set of 4 polarization curves, denoted as Horizontal (H), Vertical (V), Diagonal (D) and Anti-diagonal (A) are recorded, as shown in Fig.~\ref{fig:appendix1}. It can be observed how the payload is able to maintain high quality entanglement for different temperatures and environments, from laboratory to in-orbit operation, going through the Acceptance Test in between. It should be noted that the varying coincidence rates among experimental runs are due to the different laser currents utilized (which depend upon the environmental temperature). Different laser currents yield different laser modes, which inherently produce different coincidence rates.

From Fig.~\ref{fig:appendix1}(a), the visibilities recorded before the Acceptance Test at an average temperature of \SI{20}{\degree C} were:
$V_H$ = \SI[separate-uncertainty = true]{0.97(5)}{},
$V_V$ = \SI[separate-uncertainty = true]{0.97(6)}{},
$V_D$ = \SI[separate-uncertainty = true]{0.84(5)}{} and
$V_A$ = \SI[separate-uncertainty = true]{0.90(5)}{}, 
and the corresponding CHSH parameter was \SI[separate-uncertainty = true]{2.63(7)}{}. Fig.~\ref{fig:appendix1}(b) presents the polarization curves obtained after the Acceptance Test, and the visibilities recorded at an average temperature of \SI{24}{\degree C} were:
$V_H$ = \SI[separate-uncertainty = true]{0.95(5)}{},
$V_V$ = \SI[separate-uncertainty = true]{1.00(6)}{},
$V_D$ = \SI[separate-uncertainty = true]{0.89(5)}{} and 
$V_A$ = \SI[separate-uncertainty = true]{0.88(6)}{}.
The corresponding CHSH parameter was \SI[separate-uncertainty = true]{2.60(6)}{}.

After SpooQy-1 was launched into orbit, the highest recorded visibilities were obtained from a satellite pass on 16th July 2019 and the values were:
$V_H$ =\SI[separate-uncertainty = true]{0.98(5)}{},
$V_V$ =\SI[separate-uncertainty = true]{0.97(6)}{},
$V_D$ =\SI[separate-uncertainty = true]{0.88(6)}{} and 
$V_A$ =\SI[separate-uncertainty = true]{0.88(6)}{}.
From these values, a CHSH parameter of \SI[separate-uncertainty = true]{2.60(7)}{} was extracted. 
The polarization correlation curves obtained from this pass were presented in Fig.~\ref{fig:appendix1}(c) and the experiment was run at an average temperature of \SI{17.5}{\degree C}.

\end{appendices}
\end{document}